\documentclass[%
reprint,
superscriptaddress,
amsmath,amssymb,
aps,
pra,
]{revtex4-2}

\usepackage{url}
\usepackage[sort&compress]{natbib}
\usepackage{graphicx}
\usepackage{dcolumn}
\usepackage{bm}

\usepackage[english]{babel} 
\usepackage{xcolor} 

\begin{document}
\selectlanguage{english} 
\preprint{APS/123-QED}

\title{Long-Range Dipole-Dipole Interactions Enabled with Guided Plasmons of Matched Nanoparticle-on-Mirror Antenna Pairs}

\author{Bowen Kang}
\affiliation{School of Physics and Information Technology, Shaanxi Normal University, Xi'an, 710119, China}
\author{Huatian Hu}
\email{huatian.hu@iit.it}
\affiliation{Center for Biomolecular Nanotechnologies, Istituto Italiano di Tecnologia, Arnesano, 73010, Italy}
\author{Huan Chen}
\affiliation{School of Physics and Information Technology, Shaanxi Normal University, Xi'an, 710119, China}
\author{Zhenglong Zhang}
\email{zlzhang@snnu.edu.cn}
\affiliation{School of Physics and Information Technology, Shaanxi Normal University, Xi'an, 710119, China}

\date{\today}

\begin{abstract}
Ruling a wide range of phenomena, dipole-dipole interactions (DDI) are typically constrained to the short range due to their rapid decay with the increasing dipole separations, limiting the performance in long-range applications. By judiciously designing the photonic structures that control the two-point Green’s functions of the electromagnetic environment, the spontaneous emission of quantum emitters (luminescence) and their interactions (e.g., Förster energy transfer) can be conveniently tuned. In this paper, we designed a matched nanoparticle-on-mirror antenna pair with enhanced DDI guided by surface plasmon polaritons confined to the metal substrate, which ensures concentrated and enhanced interaction over long ranges of tens of wavelengths. The long-range ($\sim 10 \lambda$) DDI between donor-acceptor emitters is enhanced by $6\times 10^{3}$ times respective to bare gold film, and $4.4\times 10^{4}$ times respective to vacuum. Our result provides a promising testbed for investigating long-range DDI phenomena on the nanoscale. 
\end{abstract}

\maketitle
\section{Introduction}
Dipole-dipole interactions (DDI) are fundamental ingredients in a wide range of quantum phenomena involving two separate dipoles (or two-level systems), such as Förster resonance energy transfer (FRET), collective lamb shifts, and Casimir effects \cite{Lukas}. However, owing to the rapidly decaying Coulomb interactions in the near ($\sim1/r^3$) and far ($\sim1/r$) fields \cite{HMM}, DDI is typically confined to the dimension with the dipole separation much shorter than the wavelength. 

Fortunately, as one kind of electromagnetic interaction, DDI can be modulated by judiciously designing the photonic structures. For example, the energy transfer rate between the donor and acceptor dipoles in the well-known FRET is determined by both local and nonlocal electromagnetic environments depicted by two-point Green's function $\bar{\bar{\bf G}}(\textbf{r},\textbf{r’})$. The Green's function at the same position (local, $\textbf{r} = \textbf{r’}$) can describe the local density of states (LDOS), which determines the enhancement of emission and absorption rates. Meanwhile, its nonlocal propagator counterpart ($\textbf{r} \neq \textbf{r’}$) controls the coupling efficiency between them. The control of the local density of state and Purcell factor has been widely studied in the context of enhanced fluorescence \cite{Fp,Purcell1946}, cathodoluminescence \cite{Kociak2014}, and electroluminescence \cite{Baldo1999}, whereas the design of the Green's function at a remote location for controlling DDI relatively needs more elaboration \cite{ACS2018strongcoupling}. Enhancing DDI typically requires maintaining the intensity of the light emission over distance. In this sense, directionally guiding the electromagnetic fields with low loss by optical waveguides \cite{martin2011dissipation,PCW,FET}, hyperbolic metamaterials (HMM) \cite{HMM,NC,HMgoldnanorods,biehs2016long}, and plasmonic lattice \cite{nanoLetterlattice,PRLlattice} can be relevant. On the other hand, enhancing the interaction by confining the light in microcavities \cite{antolinez2019defect,andrew2000forster}, plasmonic nanocavities \cite{nanocavity,hamza2023long,hamza2021forster} for higher energy density is another strategy. These structures can generate propagating or local electromagnetic modes, facilitating longer-range or higher-intensity interaction.  

In general, as two paradigms of light-matter interaction described by Green's function, a system's localization and propagation performances are usually contradictory. One needs to balance these two effects for the optimal performance of long-range applications such as FRET. For example, localized surface plasmons (LSP) with finite structures are capable of confining light into deep subwavelength mode volume with scales of nanometer cubic \cite{baumberg2019extreme} or even picometer cubic \cite{pico}. It supports prominent localized enhanced fluorescence and absorption \cite{Lukas}, which are significant ingredients for FRET. The intensity of DDI can indeed be significantly enhanced by using metal nanoparticles \cite{hsu2017plasmon,ding2017plasmon}, yet with a very short interaction range ($\ll \lambda$) due to the deep-subwavelength confinement of the surface plasmons. The plasmons on a metallic structure with at least one extended dimension can propagate on the surface with a large \textit{k} vector and confined fields (i.e., surface plasmon polaritons (SPP)), which enables efficient remote excitation \cite{hu2022Nanophotonics,NWNCNanoletters}, and energy transfer \cite{spp} to a remote entity. However, the confinement and LDOS could be much lower than LSP. In this context, hybrid structures that can combine outstanding localized enhancement and effective coupling through guided modes are relevant. By carefully designing two distinct manifests of the plasmons, i.e., LSP and SPP, one can anticipate an efficient FRET in hybrid plasmonic structures with interaction distance being several times the wavelengths ($\geq \lambda$) \cite{ACS2018strongcoupling,spp,bouchet2016long}.

In this letter, we adopt matched nanocube-on-mirror (NCoM) pairs bridged by SPPs on the metal film to demonstrate a long-range DDI on the scale of several wavelengths. NCoMs have been widely studied due to their simplicity in fabrication, controllable structural parameters, and excellent local field enhancement effect, which results in an enhanced spontaneous emission rate \cite{NPoM}. In contrast to a single nano-optical antenna that exclusively supports LSP, NCoM pairs exhibit dual functionality by supporting both LSP and SPP. The energy conversion into SPPs and free space suggests a balance between the localized and propagating nature of the field \cite{hu2022Nanophotonics}. This work demonstrates that the DDI distance can effectively propagate over ten times the wavelengths. The DDI assisted by matched NCoM pairs at 8 $\mathrm{\mu m}$ shows a more significant enhancement of more than 3 orders of magnitude compared with bare metal surface and air. Comparisons with different structures from the literature, e.g., photonic crystal waveguides (PCW) \cite{PCW}, plasmonic lattices \cite{nanoLetterlattice}, HMM \cite{HMM}, gold film, etc., have been conducted to evaluate our current design.

\section{Theory and Methods}
FRET occurs through the dipole coupling between donor and acceptor via emitting virtual photons between molecules \cite{jones2019resonance,andrews2007principles}. 
The energy transfer is characterized by the spontaneous emission of a donor emitter towards a specific location on the acceptor. It is determined by near-field DDI ($V_{\mathrm{dd}}$) and the properties of the quantum emitters, including emission spectra of the donor emitter $\sigma_{\mathrm{D}}(\omega)$ and the absorption cross-sections of the acceptor emitter $\sigma_{\mathrm{A}}(\omega)$. 
In general, the transfer rate can be defined as \cite{RET1999}, 
\begin{equation}
\Gamma_{\mathrm{ET}}=\frac{2 \pi}{\hbar^{2}} \int d \omega\left|V_{\mathrm{dd}}(\omega)\right|^{2} \sigma_{\mathrm{D}}(\omega) \sigma_{\mathrm{A}}(\omega).
\label{eq:fret}
\end{equation}
This relation reveals the important role of Green's function at the local and remote locations. The spectral densities ($\sigma_{\rm A,D}$) are related to the intrinsic properties of the emitters as well as the local enhancement determined by Green’s function locally \cite{T2013Weak}. On the other hand, the $V_{\mathrm{dd}}$ term depends on the propagation behavior of the Green's function at a remote location.
\subsection{Two-point Green's function}
In electromagnetic theory, the dyadic Green’s function $\bar{\bar{\bf G}}\left(\mathbf{r}_\mathrm{A}, \mathbf{r}_\mathrm{D}\right)$determines the electric field at acceptor location $\mathbf{r}_\mathrm{A}$ generated by a point source at a donor location $\mathbf{r}_\mathrm{D}$ with dipole moment $\boldsymbol{\mu}_\mathrm{D}$, following the relation of $\boldsymbol{E}\left(\mathbf{r}_\mathrm{A}\right)=(\omega^{2}/\varepsilon_{0} \varepsilon_\mathrm{r} c^{2}) \bar{\bar{\bf G}}\left(\mathbf{r}_\mathrm{A}, \mathbf{r}_\mathrm{D}\right) \cdot \boldsymbol{\mu}_\mathrm{D}$ \cite{Lukas}.

The dyadic Green's function $\bar{\bar{\bf G}}$ is a symmetric $3\times3$ tensor where each element is determined by extracting the corresponding electric field components associated with a specific alignment of the electric dipole. For example, the most prominent component in an NCoM system, $G_\mathrm{zz}=(\varepsilon_{0} \varepsilon_\mathrm{r} c^{2}/\mu_\mathrm{D} \omega^{2}) E_\mathrm{z}$, can be determined by the z-component of the electric field excited by a z-polarized dipole \cite{Lukas,Green}.
The other components of the tensor could be calculated likewise. In this paper, we used the finite-element method to calculate the Green's tensor (see Methods in Appendix).


\subsection{Purcell factor and DDI}
The spontaneous emission rate depends on the density of photonic modes of the environment quantified by the partial LDOS \cite{Lukas} relying on the Green's function at the same location: $\rho\left({\bf r}_\mathrm{D}; \omega\right) = (6\omega/\pi c^{2}) \mathbf{n}_\mathrm{D} \cdot \operatorname{Im}\left\{\bar{\bar{\bf G}}\left(\mathbf{r}_\mathrm{D}, \mathbf{r}_\mathrm{D}; \omega\right)\right\} \cdot \mathbf{n}_\mathrm{D}$.
Here, we assume that the emission is from a donor with direction $\mathbf{n}_{\rm D }$ positioned at $\mathbf{r}_\mathrm{D}$. With dipole approximation, the donor emitter can be described by a dipole with moment $\boldsymbol{\mu}_\mathrm{D}= \mu_\mathrm{D}\mathbf{n}_\mathrm{D}$. Therefore, the spontaneous emission rate of a point source ($\mathbf{r}_\mathrm{D}$) at frequency $\omega$ can be calculated through the Green's function as $\gamma_{\rm D}=(\pi \omega_{0}/3 \hbar \varepsilon_{0})|\mu_\mathrm{D}|^{2} \rho\left({\bf r}_\mathrm{D}; \omega\right)=(2 \omega_\mathrm{D}^{2}\left|\mathbf{\mu}_\mathrm{D}\right|^{2}/\hbar \epsilon_{0} c^{2})\left[\mathbf{n}_\mathrm{D} \cdot \operatorname{Im}\left\{\bar{\bar{\bf G}}\left(\mathbf{r}_\mathrm{D}, \mathbf{r}_\mathrm{D}; \omega\right)\right\} \cdot \mathbf{n}_\mathrm{D}\right]$.

The Purcell factor $F_\mathrm{p}$ can be characterized by the enhancement of the spontaneous emission rate which contains both radiative and non-radiative components, as determined by the LDOS $\rho\left(\mathbf{r}_\mathbf{D};\omega\right)$ compared to that in vacuum $\rho^{0}\left(\mathbf{r}_\mathbf{D};\omega\right)$ \cite{2006Enhancement,chen2022sub}:
\begin{equation}
{F}_\mathrm{p} =\frac{\gamma_{\mathrm{D}}}{\gamma_{\mathrm{D}}^{0}}=\frac{\rho\left(\mathbf{r}_\mathrm{D}; \omega\right)}{\rho^{0}\left(\mathbf{r}_\mathrm{D} ; \omega\right)}.
\label{eq:Fp}
\end{equation}
The vacuum LDOS reads, $\rho^{0}\left(\mathbf{r}_\mathrm{D} ; \omega\right)=\frac{\omega^{2}}{\pi^{2} c^{3}}$ \cite{Lukas}. On the other hand, the resonant DDI potential for two dipoles in an arbitrary photonic environment is also governed by Green's function \cite{OE}:
\begin{equation}
V_\mathrm{dd}\left(\mathbf{r}_\mathrm{A},\mathbf{r}_\mathrm{D};\omega\right) = -\frac{\omega^2}{\varepsilon_{0}c^2}\boldsymbol{\mu}_\mathrm{A}\cdot\bar{\bar{\bf G}}\left(\mathbf{r}_\mathrm{A},\mathbf{r}_\mathrm{D};\omega\right)\cdot \boldsymbol{\mu}_\mathrm{D},
\label{eq:Vdd}
\end{equation}
where $\boldsymbol{\mu}_\mathrm{D}$ and $\boldsymbol{\mu}_\mathrm{A}$ are the dipole moments of the donor and acceptor emitters, respectively. $\varepsilon_{0}$ is the vacuum permittivity, and $c$ is the speed of light.

The enhancement of the DDI interaction can be quantified by a factor ${n}_\mathrm{ET}$, which normalizes the values by themselves in the vacuum. This factor describes the efficiency of long-range energy transfer between the donor and acceptor \cite{FET}:

\begin{equation}
{n}_\mathrm{ET}=\frac{\left|V_\mathrm{dd}\left(\mathbf{r}_\mathrm{A},\mathbf{r}_\mathrm{D};\omega\right)\right|^2}{\left|V^{0}_\mathrm{dd}\left(\mathbf{r}_\mathrm{A},\mathbf{r}_\mathrm{D};\omega\right)\right|^2} .
\label{eq:net}
\end{equation}

Furthermore, as a figure of merit, we can define an enhancement factor $F^*$ for the energy transfer rate $\Gamma_{\rm ET}$ in Eq.\eqref{eq:fret}. This factor is defined as the product of the Purcell factor (${F}_\mathrm{p}$) and the energy transfer rate (${n}_\mathrm{ET}$), effectively capturing both the enhancement of the LDOS and the efficiency of DDI, as expressed:
\begin{equation}
{F}^{\ast}={F}_\mathrm{p}\times{n}_\mathrm{ET}.
\label{eq:F*}
\end{equation}
\begin{figure*}
\includegraphics[width=0.8\linewidth]{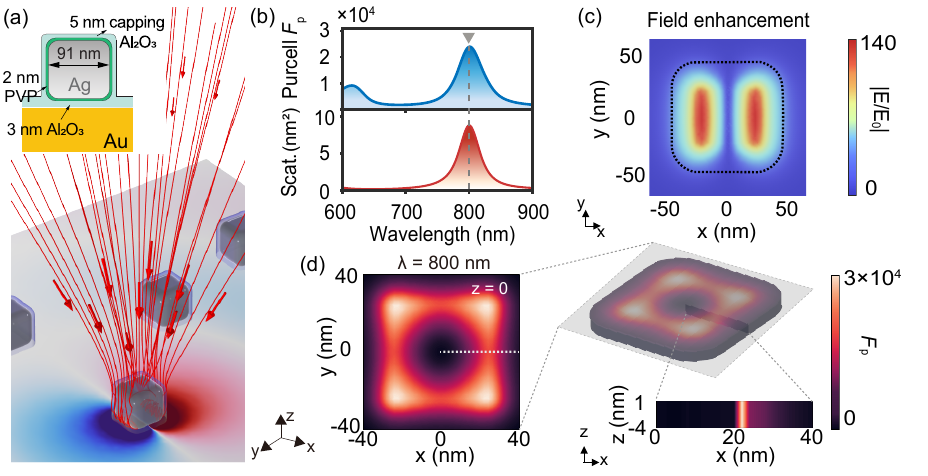}
\caption{\label{fig1}\textbf{NCoM antenna harvests and guides light.} (a) Light harvesting is inferred by red streamlines of the Poynting vector with arrows. The SPPs generated on the metal film are indicated by the colormap on the substrate. The inset demonstrates the cross section of the antenna. (b) Simulated Purcell factor and scattering spectra of NCoM. (c) Electric field enhancement distributions of the cavity plasmon modes in the Donor NCoM systems under an 800 nm (on resonance) linearly polarized plane-wave excitation. (d) The distribution of the Purcell factor at $\lambda=800$ nm under the nanocube in the gap. It depends on the imaginary part of the Green's function at the same positions.}
\end{figure*}
Here in Eq. \ref{eq:F*}, for simplicity, we use the Purcell factor to qualitatively indicate the enhancement of the cross-sections shown in Eq. \ref{eq:fret}. More specifically, one can also consider the quantum efficiency of the antennas for further comparison.

To our knowledge, we have chosen one of the best types of nanoantennas for enhancing light harvesting (enhanced absorption) and emission (enhanced fluorescence) with an outstanding quantum efficiency \cite{baumberg2019extreme,maikenmikkelsen}. Meanwhile, it can convert an estimated one-third of the energy contained by the modes into SPPs, ensuring efficient SPP generation which is important for enhancing $V_{\rm dd}$ \cite{hu2022Nanophotonics}. As required by FRET, this structure can balance both the localization and propagation of the mode. In this context, this paper will concentrate more on investigating and optimizing the interaction term $V_{\mathrm{dd}}$ for energy transfer by evaluating the dyadic Green's function at the acceptor emitter location $\mathbf{r}_{\rm A}$, due to the presence of the donor emitter at $\mathbf{r}_\mathrm{D}$. 

\section{Results and discussion}
Figure \ref{fig1}(a) shows the schematic of the proposed structure of the NCoM antenna pairs. By drop-casting silver nanocubes on a gold metal film, one can easily obtain antenna pairs by selecting two nanocubes with proper distance on the substrate \cite{hu2022Nanophotonics}. We start by analyzing the receiving antenna (Fig. \ref{fig1}(a)) which will couple with the donor emitter. As shown by the inset of Fig. \ref{fig1}(a), it consists of a 91 nm diameter silver nanocube covered with 2 nm PVP deposited on a gold film with 3 nm Al$_2$O$_3$ nanogaps as a spacer and 5 nm Al$_2$O$_3$ coating as a protecting layer. Therefore, the overall gap thickness (PVP and ALD spacer) between silver nanocube and gold film is 5 nm, which will not introduce significant nonlocal effect that may shift the energy and decrease the enhancement \cite{2012Probing}. The refractive index of the gold and silver are taken from the work of Johnson et.al, \cite{johnson1972optical}. Previous studies have demonstrated that compared to the nanosphere-on-mirror (NSoM) and nanowire-on-mirror, the NCoM exhibits superior scattering and SPP efficiencies \cite{hu2022Nanophotonics}, thereby enabling effective control of plasmonic near fields for enhancing DDI. The streamlines in Fig. \ref{fig1}(a) infer how well the antenna can harvest the light from the free space. The color map on the surface represents the excited SPPs on the metal. In Fig. \ref{fig1}(b), the normalized scattering spectrum excited by a normal incident light shows that the magnetic dipole mode of the nanopatch antenna situates at $\lambda$ = 800 nm. Due to the small nanogap between the cube and the metal film, the electric field can be enhanced 140 times at the resonance (Fig. \ref{fig1}(c)). This electric field enhancement will give rise to an intensity enhancement factor of $|\textbf{E}|^2/|\textbf{E}_0|^2$=140$^2$ acting on the spectral cross-section $\sigma_{\rm A,D}$.

Due to the reciprocity, the Purcell factor is also greatly enhanced by the NCoM nanopatch antenna. As shown in the upper panel of Fig. \ref{fig1}(b) and the mapping in Fig. \ref{fig1}(d), the Purcell factor of a dipole situated at the corner of the nanocube also shows a prominent resonance at 800 nm with a $3\times10^4$ times of enhancement. The Purcell factor is higher when the dipole is positioned at a corner of a cube, as compared to central or edge positions. This feature serves as the basis for optimizing NCoMs pair design parameters to enhance long-range DDI.  

\begin{figure}
\centering\includegraphics[width=7.5cm]{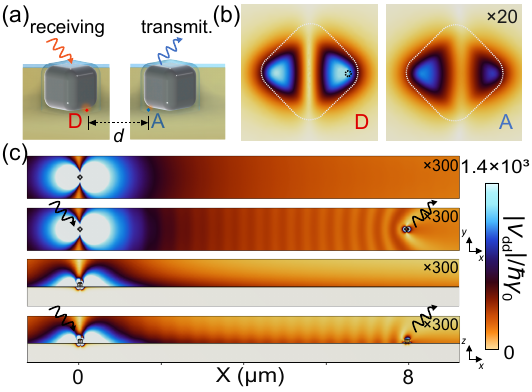}
\caption{\label{fig2}\textbf{Energy transfer between donor-acceptor emitters resided in nanoantenna pairs.} (a) Schematic of the DDI enhanced by NCoM-NCoM antenna pairs. Here, `$d$' denotes the distance between interacting emitters, `D' represents the donor emitter, and `A' represents the acceptor emitter. The distribution of ${\left |V_\mathrm{dd}\right | /\hbar \gamma _{0}}$ by the dipole in donor NCoM and the NCoM-NCoM pairs system with a distance of 8 $\mathrm{\mu m}$ (10$\lambda$), (b) zoom-in figure of the fields under the cubes, (c) the overall distribution between two antennas containing donor and acceptor respectively.}
\end{figure}

When we have two identical nanopatch antennas separated at a distance of $d$ as shown in Fig. \ref{fig2}(a), we obtain an antenna pair. Because of the identical size, they have matched resonant energies and the SPP excited from one antenna can efficiently drive another. Through this guided plasmon mode, two antennas are linked together. Experiments have seen the SPPs from one nanoantenna can be strong enough to remotely excite the Raman and photoluminescence spectra from the other antenna 10 $\mu$m away \cite{hu2022Nanophotonics,NWNCNanoletters}.
For quantitatively evaluating the DDI, we implement a dimensionless value to describe the intensity of the DDI as $|V_\mathrm{dd}|/\hbar \gamma_0$ \cite{HMM}, where the denominator represents the energy of dipole decay rate in the vacuum. Figs. \ref{fig2}(b) and (c) show the DDI intensity distribution for an NCoM pair with a donor-acceptor separation of 8 $\mathrm{\mu m}$ (10$\lambda$). The nanocube is assumed to be rotated to 45 degrees and the emitter dipole is positioned at the corner (marked by the circle in Fig. \ref{fig2}(b)), as this is the most efficient way of generating SPP and obtaining a high LDOS as well as $V_{\rm dd}$. 
In the map of DDI $|V_\mathrm{dd}|/\hbar \gamma_0$ (Fig. \ref{fig2}(c)), we can clearly see interference fringes caused by the launched SPPs from the receiving antenna (donor) and the bounced-back reflection from the transmittance antenna (acceptor). At a remote location with an 8 $\mu$m distance (``A" of Fig. \ref{fig2}(b)), the $|V_\mathrm{dd}|/\hbar \gamma_0$ drops about 25 times in magnitude, compared with that in the receiving antenna where the donor is situated (``D" of Fig. \ref{fig2}(b)). 

\begin{figure}[!h]
\centering\includegraphics[width=7.5cm]{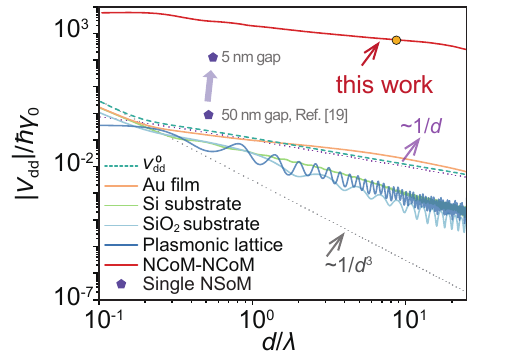}
\caption{\label{fig3}\textbf{Comparison of interaction between antenna pairs and the various photonic environments.} The ${\left|V_\mathrm{dd}\right|/\hbar \gamma _{0}}$ between emitters in various photonic environments. $d/\lambda$ is the effective interaction distance normalized to the emission wavelength.}
\end{figure}

\begin{figure*}[!ht]
\centering\includegraphics[width=0.8\linewidth]{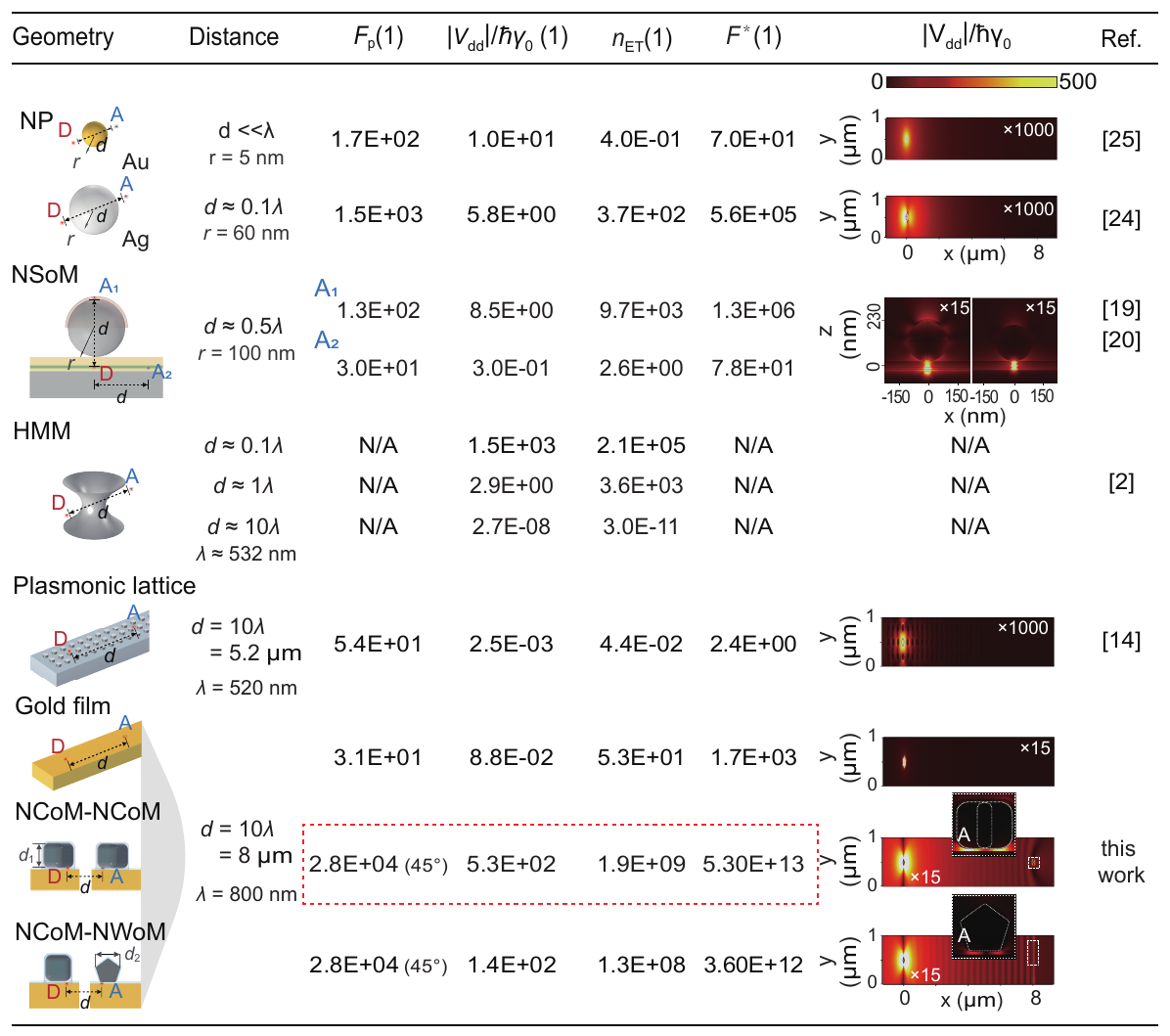}
\caption{\label{table}Comparison of ${F}_\mathrm{p}$, ${\left|V_\mathrm{dd}\right|/\hbar \gamma _{0}}$, ${n}_\mathrm{ET}$} and the figure of merit $F^*$ performances of different systems, including nanoparticles (Au \cite{ding2017plasmon}, Ag \cite{hsu2017plasmon}), NSoM \cite{hamza2023long,hamza2021forster}, HMM \cite{HMM}, plasmonic lattice \cite{nanoLetterlattice}, dipole on the bare gold film, NCoM-NCoM (corner to corner at 45°), and NCoM-NWoM (45°). All maps share the same colorbar.
\end{figure*}

To validate the advantages of the NCoM-NCoM pairs system, we calculated and compared the interaction potential ${\left|V_\mathrm{dd}\right|/\hbar \gamma _{0}}$ in different systems including bulk substrates with different materials and plasmonic lattices \cite{nanoLetterlattice}) in Fig. \ref{fig3}. Beyond that, in Fig. \ref{table}, we also summarize the values of Purcell factor ${F}_\mathrm{p}$, ${\left|V_\mathrm{dd}\right|/\hbar \gamma _{0}}$, ${n}_\mathrm{ET}$ and the figure of merit $F^*$ of different structures including those from literature \cite{ding2017plasmon,hsu2017plasmon,hamza2023long,hamza2021forster,HMM,nanoLetterlattice}. Noticeably, the ${\left |V_\mathrm{dd}\right | /\hbar \gamma _{0}}$ map between the donor and acceptor for each system is drawn with a separation of $\simeq10\lambda$, with a magnified view specifically at the receptor location as illustrated in the insets of Fig. \ref{table}. In general, when compared to simple substrates such as metal and dielectric (Fig. \ref{fig3}), NCoM antenna pairs can have a much higher interaction due to the enhancement of each nanoantenna. Interestingly, a recent experiment of a single NSoM antenna, with dipoles separated by roughly $\sim\lambda/2$, with a large gap thickness of 50 nm \cite{hamza2023long}, has validated the effectiveness of the DDI in nanogap antenna systems. As indicated by the purple markers in Fig. \ref{fig3}, it has a stronger DDI compared to various configurations, and by decreasing the gap thickness to 5 nm like this work, the ${\left |V_\mathrm{dd}\right | /\hbar \gamma _{0}}$ of the single NSoM could be further enhanced. When compared to the waveguide-like structures with low losses, NCoM pairs with SPPs between them will have higher decay. But when considering an interaction with a separation of about 10$\lambda$, NCoM pairs (red curve of Fig. \ref{fig3}) will still be at the sweet point and have higher interaction intensity than other structures due to the double enhancements and the guided plasmons.


More detailed comparisons with the structures presented in the previous literature are shown in Fig. \ref{table}. First of all, two dipoles near the surface of the gold  \cite{ding2017plasmon} and silver \cite{hsu2017plasmon} nanoparticles can have prominent Purcell factor as well as DDI due to the near field enhancement of the LSP. Literature \cite{hsu2017plasmon,ding2017plasmon} shows a DDI from two dipoles separated by 130 nm and 20 nm with a 120 nm silver and 10 nm gold nanoparticle in between, respectively. In this case, the dipoles are considered at different sides of the nanoparticle, and the separation of the dipole to the surface of the nanoparticles is only 5 nm each. Therefore, this DDI will rely on the help of the nanoparticle which will highly depend on the particle's size. When the distance between the two dipoles becomes larger, there is a sharp drop in the DDI. To improve the performance of a single nanoparticle, an NSoM configuration \cite{hamza2023long,hamza2021forster} with a nanogap can experimentally achieve effective DDI with a distance on the order of half of the wavelength in a single nanocavity due to the confined mode volume. Ones have reasons to believe that the single NSoM antenna if well matched into pairs as the design in Fig. \ref{fig2} suggests, can ideally have similar remote interaction performance but with slightly lower excitation efficiency \cite{hu2022Nanophotonics}.

On the other hand, the long-range effects of DDI have also been experimentally confirmed in systems such as plasmonic lattice \cite{nanoLetterlattice}, HMM \cite{HMM}, and PCW \cite{PCW}. In these systems, the range of DDI is able to extend to the distance that is comparable with or even larger than the wavelength $d/\lambda > 1$. HMM as a kind of newly-emerged low-loss metamaterial that can support a very high ${\left|V_\mathrm{dd}\right|/\hbar \gamma _{0}}$ on the order of $0.1\lambda$ yet may decay rapidly following a $d^{-3}$ dependence with the increasing distance \cite{HMM}. Low-loss PCW \cite{PCW} is a good structure to concentrate and guide the waveguide mode to keep an effective coupling in the long range. The plasmonic lattice mode can have a greater Purcell factor thanks to the resonance of the LSP to support relatively high figures of merit $F^*$ at a long range at $10\lambda$ despite the plasmonic loss.

When considering a metallic film as substrate and further constructing a nanoantenna pair system on top as we did, the $F_{\rm p}$, ${\left|V_\mathrm{dd}\right|/\hbar \gamma _{0}}$, as well as the $F^*$ could be significantly enhanced.
The signal from the donor dipole is first amplified by the receiving antenna with enhanced LDOS, then transferred through guided SPP on the metallic film to the transmitting antenna, at which the energy will be amplified again and transferred to the acceptor emitter. Taking different substrates including SiO$_{2}$, Si, and Au film as control groups, the results are shown in Figs. \ref{fig3} and \ref{table} demonstrate that NCoM-NCoM nanoantenna pairs exhibit significantly enhanced DDI, surpassing conventional systems by 3 orders of magnitude. Especially when considering the energy transfer rate enhancement $F^*$, the proposed NCoM antenna pair has shown a well-balanced trade-off between enhancing spontaneous emission decay rates and DDI, manifesting a striking figure of merit $F^*$ over long distances over the conventional configurations. 

\section{Conclusion}
In summary, the introduction of matched nanoparticle-on-mirror antenna pairs significantly enhances long-range DDI by optimizing both local and remote electromagnetic environments. Our findings show that these plasmonic nanoantenna pairs amplify the DDI strength between donor-acceptor emitters at separation distances exceeding $\sim 10 \lambda$ by 3 orders of magnitude compared to HMM, plasmonic lattices, bare gold films, etc. When considering an overall figure of merit $F^*$ at a long distance, our nanoantenna pairs can even have more than 8 orders of magnitude of improvement. This design provides a tool for long-range energy transfer applications and serves as a platform for studying long-range DDI at the nanoscale.

\section{Author Contributions}
H.H. conceived the idea and supervised the calculations. B.K. performed the theoretical calculations and created the figures. H.H., B.K., H.C. and Z.Z. wrote the manuscript. Z.Z. and H.H. supervised the project. All authors discussed and commented on the manuscript.

\section{Conflict of Interest}
The authors declare no conflict of interest.

\begin{acknowledgments}

This work was supported by the National Key R\&D Program of China (No. 2024YFA1409900), the National Natural Science Foundation of China (Nos. 12474389, U22A6005, 12204362, and 12304426), the Natural Science Foundation of Shaanxi Province (No. 2024JC-JCQN-07), the Fundamental Science Foundation of Shaanxi (No. 22JSZ010), and the Fundamental Research Funds for Central Universities (Nos. GK202308001, LHRCTS23063).
\end{acknowledgments}

\appendix
\section{Influence of material parameters on DDI strength}
While plasmon can indeed enhance near-field interactions and transmit information across wavelength-scale distances, the trade-off with losses remains critical. For practical applications, addressing these loss issues is essential. In our structural optimization, we have chosen metal materials (such as silver or gold) with lower absorption and scattering losses. Additionally, to avoid maybe overly optimistic assumptions (as in the case of Johnson \& Christy’s parameters), we have compared refractive index data from different researchers for gold. We also evaluated the DDI of the NCoM-NCoM pairs system at a relative angle of 45°, as shown in Table \ref{tab:table1}. Our findings indicate excellent enhancement of DDI despite varying conditions. 

\begin{table}[!ht]
\caption{\label{tab:table1} For comparison, the ${F}_\mathrm{p}$, ${\left|V_\mathrm{dd}\right|/\hbar \gamma _{0}}$ and $F^*$ performances of NCoM-NCoM systems (corner to corner at 45°) were calculated using data from Olmon et al. \cite{olmon2012optical}, Rakic et al. \cite{rakic1998optical}, Werner et al. \cite{werner2009optical}, Ordal et al. \cite{ordal1987optical}, and Babar and Weaver \cite{babar2015optical}.}
\begin{ruledtabular}
\begin{tabular}{cccc}
\textrm{Data}&
\textrm{${F}_\mathrm{p}$}&
\textrm{${\left|V_\mathrm{dd}\right|/\hbar \gamma _{0}}$}&
\textrm{$F^*$}\\
\colrule
Olmon et al.& 2.4E+04& 4.17E+02& 2.9E+13\\
Rakic et al. & 1.4E+04 & 2.4E+02 & 5.7E+12\\
Werner et al.& 1.6E+04 & 1.8E+02 & 3.4E+12\\
Ordal et al. & 2.3E+04 & 3.4E+02 & 1.8E+13\\
Babar and Weaver& 3.1E+04 & 7.7E+02 & 1.2E+14\\
\end{tabular}
\end{ruledtabular}
\end{table}
\section{The influence of the dipole moment orientation on DDI strength}
\begin{figure}[!ht]
\centering\includegraphics[width=7cm]{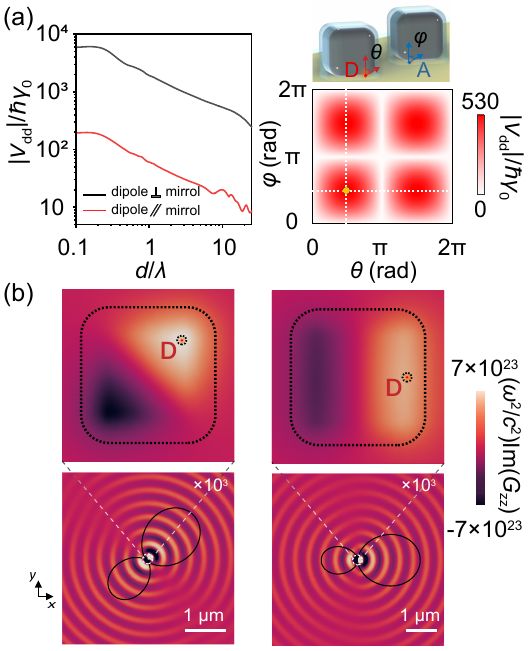}
\caption{\label{fig5} (a) The calculated DDI ${\left |V_\mathrm{dd}\right |/\hbar\gamma _{0}}$ between donor-acceptor emitters pair with different orientations and varying distances. Right panel: Donor-dependent orientation in the NCoM-NCoM pairs system with $d$ = 8 $\mathrm{\mu m}$. Schematic representation of the angular orientations of donor and acceptor dipoles (inset) and (b) directional radiation effects in NCoM. Black curves are the far-field radiation direction.} 
\end{figure}

In Fig. \ref{fig5}, the inset illustrates the angular orientations of the donor and acceptor dipoles. Specifically, $\theta$ and $\varphi$ represent the polar angles of the emitters under two antennas, respectively. These orientations play a crucial role in determining the intensity of the DDI. When the donor’s dipole moment aligns perpendicularly to the mirror, the DDI intensity is significantly higher compared to when it aligns parallel to the mirror. This observation underscores the importance of dipole orientation in shaping the DDI. Furthermore, Fig. \ref{fig5}(b) reveals intriguing directional SPP radiation from NCoM. The SPP propagates along the metallic surface when the dipoles are located at the corner and edge center of the cube. When positioned at a corner, the radiation direction aligns along the diagonals. On the other hand, when located on the sides of the cube, the radiation is perpendicular to the cube. By strategically adjusting the position of the dipole within the gap between the cube and substrate, one can precisely control the SPP propagation direction. These findings offer valuable insights for designing nanophotonic devices and manipulating light-matter interactions. 

\section{Methods}
In the full-wave simulation package COMSOL Multiphysics 6.1, a three-dimensional geometric structure was established, with the dielectric function of metal taken from Johnson et al. \cite{johnson1972optical}. The refractive index of PVP and Al$_{2}$O$_{3}$ are set to be 1.5. Perfectly matched-layer (PML) were applied as boundary conditions. Following the same methods presented in our previous work \cite{hu2022Nanophotonics,chen2022sub}, when calculating the scattering problems from the NCoM, a standard two-step model was used where the first step calculated the background field without the nanocube whereas the second step calculated the scattering from the nanocube. Secondly, the dyadic Green's function, $\bar{\bar{\bf G}}$, as defined by the electric field at position $\mathbf{r}_\mathrm{A}$ generated by a point source at position $\mathbf{r}_\mathrm{D}$ with dipole moment $\boldsymbol{\mu}_\mathrm{D}$. Subsequently, the components of the dyadic Green's function were calculated by running simulations with the test dipoles oriented along the x-, y-, and z-directions, where the corresponding components of the electric fields are investigated \cite{Green}. As illustrated in Ref. \cite{Lukas,Green}, the derived Green's function allows us to obtain the rate of energy dissipation, LDOS, and the spontaneous emission rate. Consequently, we derive the Purcell factor, $V_\mathrm{dd}$, $n_\mathrm{ET}$, and the enhancement factor $F^*$.

\bibliography{reference}

\end{document}